\documentclass[twoside]{article}
\usepackage{epsfig.sty}
\newcommand{\beq}{\begin{eqnarray}}
\newcommand{\eeq}{\end{eqnarray}}

\begin{document}
\title{Charmed Baryon Decay to a Strange Baryon Plus a Pion Using QCD  
Sum Rules}
\author{Leonard S. Kisslinger$^{1}$ and Bijit Singha$^{2}$\\
Department of Physics, Carnegie Mellon University, Pittsburgh, PA 15213\\
\hspace{1cm}(1) kissling@andrew.cmu.edu\hspace{1cm}(2) bsingha@andrew.cmu.edu}
\date{}
\maketitle
\begin{abstract}

This is an extension of the prediction of strange baryon decays to the
decays of charmed baryons using QCD Sum Rules. Using QCD Sum Rules we
estimate the decay $\Lambda^+_c (udc) \rightarrow \Lambda^o_s(uds)+ \pi^+$.
Although some weak decays of the $\Lambda^+_c$ have been measured, since it is 
difficult to measure $\Lambda^+_c \rightarrow \Lambda^o_s+ \pi^+$ our 
estimates should be useful for future experiments.

\end{abstract}


\noindent
PACS Indices: 12.15.Ji, 13.30.Eg, 12.38.Lg, 11.50.Li

\section{Introduction}

Many years ago the method of QCD sum rules was used\cite{hhk02}
to estimate the weak decays $\Sigma^- \rightarrow n+\pi^-, \Sigma^+ 
\rightarrow  n+\pi^+$. Using similar theoretical methods we estimate the weak
 decay  $\Lambda^+_c (udc) \rightarrow \Lambda^o_s(uds) + \pi^+$.

  The method of QCD sum rules was introduced by Shifman, Vainshtein, and 
Zakharov \cite{SVZ} to estimate properties of hadrons using 2-point 
correlators.
Recently we have estimated the masses of the charm baryon, $\Lambda^+_c$, 
bottom baryon $\Lambda^0_b$, strange baryon $\Lambda^0_s$ using QCD sum rules 
with a 2-point correlator\cite{kb17}.
We review how this method was extended to 3-point correlator by L.J. Reinders 
et. al.\cite{RRY85} which we use in the present work to estimate the rate of 
weak decays of a charm hadron to a strange hadron, similar to the weak decays
$\Sigma^- \rightarrow n+\pi^-, \Sigma^+ \rightarrow  n+\pi^+$ but with a
$charm \rightarrow strange$ rather than a $strange \rightarrow up$ quark
transition in Ref\cite{hhk02}. Since we only consider weak decays of charm to 
strange baryons, in our review we only discuss QCD sum rules using 3-point 
correlator. Recent experimental measurements of $\Lambda^+_c \rightarrow pK^- 
\pi^+$ and other $\Lambda^+_c$ decay modes\cite{BESIII16} are important but not
 directly related to the present estimate of $\Lambda^+_c \rightarrow 
\Lambda^o_s+ \pi^+$.

For the weak decay $\Lambda^+_c \rightarrow  \Lambda^o_s + \pi^+$ the weak
Hamiltonian $H_W$ is used\cite{hhk02,dgh86}. The weak decay $c \rightarrow s$,
where $s, c$ are strange, charm quarks, is needed for the calucalation of 
$\Lambda^+_c \rightarrow  \Lambda^o_s + \pi^+$.
A number of parameters for our calculation of $\Lambda^+_c 
\rightarrow  \Lambda^o_s + \pi^+$, including  $\theta_C$, the Cabibbo
angle,  are not known very well,
and we only use the main process shown in Figure 1. Although our calculation is
only an estimate of this weak decay it should be useful for future 
experiments.  

\section{QCD Sum Rule with 3-pt Correlator for Weak Decay  
$\Lambda^+_c (udc) \rightarrow  \Lambda^o_s(uds)+ \pi^+$}

For the 3-point correlator to estimate $\Lambda^+_c (udc) \rightarrow 
\Lambda^o_s(uds) \pi^+$ we need the currents for $\Lambda^+_c (udc)$ and 
$\Lambda^o_s(uds)$, and the  weak Hamiltonian. We consider here the same currents 
that we used for $\Lambda^+_c (udc)$ and $\Lambda^o(uds)$ in \cite{kb17} to 
estimate the lambda baryon masses:
\beq
\label{lambdacurrent}
\eta_{\Lambda^+_c}(x)  &=& \epsilon^{abc} [u^{aT}(x) C \gamma_\mu d^b(x)] 
\gamma^5 \gamma^\mu c^c(x) \; ,\\
\eta_{\Lambda^o_s}(x) &=& \epsilon^{abc} [u^{aT}(x) C \gamma_\nu {d}^{b}(x)]
 \gamma^5 \gamma^\nu s^c(x)  \nonumber
\eeq

\noindent
The weak Hamiltonian is 
\beq
\label{HW}
H_W &=&\frac{G_F}{\sqrt{2}}J^\mu J_\mu^\dagger\; , 
\eeq
\beq
\label{Jmu}
J^\mu &=&  V_{cs}~ \bar{s} \gamma^\mu (1-\gamma_5)c \;
+ V_{ud}~  \bar{d}\gamma^\mu(1-\gamma_5)u \;\; ,
\eeq
where $G_F$ is the Fermi coupling constant and $V_{ud}= 0.97420 \pm 0.00021$ and
$V_{cs}= 0.997 \pm 0.017$ are the elements of Cabibbo-Kobayashi-Maskawa matrix 
\cite{pdg2018}. Notice that $V_{cs}$ is close to 1, which makes this weak decay 
Cabibbo-favored. 

The QCD diagram which is used for the 3-pt correlator to estimate 
$\Lambda_c^+(udc) \rightarrow \Lambda_s^0(uds) + \pi^+$ is shown in Figure 1. 
In this diagram, the charm-strange transition and the pion creation is mediated
by a weak gauge boson, $W^+$, represented by the wavy line. There are other 
higher order diagrams corresponding to the same process but their contribution 
is negligible compared to this leading order process, so we ignore them.

\begin{figure}[ht]
\begin{center}
\epsfig{file=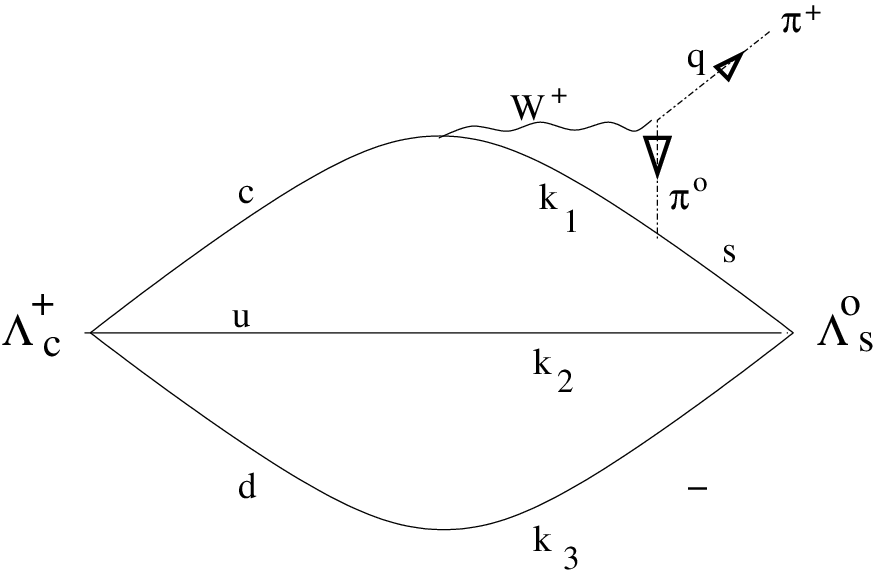,height= 6cm,width=10cm}
\end{center}
\caption{Weak Decay of $\Lambda^+_c$ to $\Lambda^o_s + \pi^+$ }
\label{Figure 1}
\end{figure}

 Note that the momentum of the strange quark is the momentum of the charm
quark minus the momentum of the pion ($k_1-q$).

The 3-pt correlator is 
\beq
\label{3pt-correlator}
\Pi_3(p,q) &=& i\int d^4x d^4y \; e^{ip \cdot x} e^{iq \cdot y}
\Pi_3(x,y) \; , 
\eeq 
where
\beq
\label{3pt-correlator-x-y}
 \Pi_3(x,y)&=&<0|T[\eta_{\Lambda^+_c}(x) H_W(y) \bar{\eta}_{\Lambda^o_s}(0)]
|0>_{\pi^+} \; .
\eeq
where the subscript $\pi^+$ denotes that the constituent quarks of the lambda 
baryons propagate in an external pion field. 
\noindent
Using $\pi^+=|u\bar{d}>, \pi^0 \simeq|d\bar{d}>$ and Eq(\ref{Jmu})\cite{hhk02},
we write the weak matrix element
\beq
\label{quark-J}
  <\pi^+|J_\alpha|\pi^0>&=& \sqrt{2}F_\pi V_{ud}~ q_\alpha \; ,
\eeq
where $F_\pi$ is the weak pion form factor and $q$ is the momentum of the pion. 
After a few lines of calculations (see Appendix A) that mainly involves trace 
identities, we obtain the following expression for the 3-point function:
\begin{eqnarray}
\Pi_{3}(p, q) &=&
-16 im_c G_F F_\pi V_{ud} V_{cs} ~  \Pi_{3Q}(p,q)
\; ,
\end{eqnarray}
where  
\begin{eqnarray}
\label{Pi3Q}
\Pi_{3Q}(p,q)
&=&
q_\nu \left[
\Pi^{\mu \mu \nu} + \Pi^{\mu \nu \mu} - \Pi^{\nu \mu \mu}
\right] , \\
\label{Pimunuomega}
\Pi^{\mu \nu \omega}(p,q)
&=&
\int \frac{d^4 k}{(2\pi)^4} \frac{k_\mu}{k^2}
\int \frac{d^4 l}{(2 \pi)^4}
\frac{l^\nu (l - k - p - q)^\omega}{l^2 \left[(l - k - p)^2 - m_c^2  \right] 
\left[  (l - k - p - q)^2 - m_s^2\right]}
\end{eqnarray}
We evaluate $\Pi^{\mu \mu \nu}$, $\Pi^{\mu \nu \mu}$ and $\Pi^{\nu \mu \mu}$ in 
Eq.~(\ref{Pi3Q}) using a dimensional regularization technique that uses 
Schwinger's proper-time representation of the propagator, $1/(p^2 - m^2) = 
\int_0^\infty d \alpha e^{-\alpha\left( p^2 - m^2 \right)}$, with the generalization of 
the gaussian integrals to $D=(4 - 2 \epsilon)$-dimensions \cite{Ashmore1973, 
Leibbrandt1975}. The detailed calculation is shown in Appendix B and we are 
mentioning the result here:
\begin{eqnarray}
\label{Pi3QEvaltd}
\Pi_{3Q}(p,q)
&=&
\frac{1}{2(4 \pi)^4}
\int_0^1 d \gamma d \rho 2 \rho^3 (1 - \rho)^2~(p.q) \int_0^1 d \kappa 
\frac{1}{g(\rho, \kappa)^3}\Big[
\kappa (p+ \gamma q)^2 \Gamma(\epsilon) a^{-\epsilon} \nonumber\\&&
\qquad \qquad + \rho \Gamma(-1 + \epsilon)a^{1 - \epsilon}\Big]
+ \frac{1}{2(4 \pi)^4} \int_0^1 d \gamma d \rho~ \frac{2p.q}{(1- \rho)}~ 
\left[ \Gamma(-1 + \epsilon)(a')^{1 - \epsilon} \right] \nonumber\\
\end{eqnarray}
where
\begin{eqnarray}
{\mathcal{Z}}^2(\gamma) &=& (1 - \gamma)m_c^2 + \gamma m_s^2, \\
g(\rho, \kappa) &=&\kappa \rho(1 - \rho)+ (1 - \kappa), \\
a(\rho, \kappa, \gamma) &=& - \kappa \rho {\mathcal{Z}}^2 + \kappa \rho 
(1 - \rho) p.(p+ \gamma q) - \Big\{ \frac{\kappa \rho (1 - \rho)}{g} \Big\} 
\kappa \rho (1 - \rho) (p+ \gamma q)^2, \\
a'(\rho, \gamma) &=& - \rho {\mathcal{Z}}^2 + \rho (1 - \rho) 
p.(p+ \gamma q) - \rho (1 - \rho) (p + \gamma q)^2 ~.
\end{eqnarray}
From Eq. (B.22) one can expect $\Pi_{3Q}(p,q)$ to have $\not{p}$ and
$\not{q}$ terms, but as Eq(97) shows these terms cancel. After carrying out
the $k$ integrals shown in Eqs(99-102) the final form for $\Pi_{3Q}(p,q)$ in
Eq(10) is shown in Eq(103).

Notice that, in Eq.~(\ref{Pi3QEvaltd}), $\Pi_{3Q}(p,q)$ will have both power and
 logarithmic divergences. The former ones are insignificant for our purpose and
 we are only be interested in log-divergences. We address this issue in
 the next section, where we apply Borel transform on Eq.~(\ref{Pi3QEvaltd}) to 
extract physical information relevant to the decay process.

\section{Borel Transformation of $\Pi_{3Q}(p,q)$}
In this section, we carry out a Borel transformation $\mathcal{B}$ on 
$\Pi_{3Q}(p,q)$ to ensure rapid convergence of the integrals: 
\begin{eqnarray}
{\mathcal{B}}_{M^2} \Pi_{3Q} (P^2)
&=&
\tilde{\Pi}_{3Q}(M^2).
\end{eqnarray}
We write $\Gamma(-1 + \epsilon) a^{1 - \epsilon} = a \ln a$ and $\Gamma(\epsilon) 
a^{- \epsilon} = - \ln a$, ignoring the power-divergent terms that vanish with 
Borel transform. This finally gives us an expression
\begin{eqnarray}
\label{Pi3cEvaltd_ZeroOne_Final_redefined}
\Pi_{3Q}(p,q) &=&
\frac{1}{2(4 \pi)^4} \int_0^1 d \gamma d \rho d \kappa \frac{2 \rho^3 
(1 - \rho)^2}{g(\rho, \kappa, \gamma)^3} (p.q) \left[
\kappa (p+ \gamma q)^2 \left( - \ln a \right) + \rho \left( a \ln a \right)
\right] \nonumber\\&& \qquad
+ \frac{1}{2(4 \pi)^4} \int_0^1 d \rho d \gamma \frac{2p.q}{(1 - \rho)} 
\left[ a' \ln a' \right]~.
\end{eqnarray}

\noindent
Notice that, we have two parameters here in our expressions ( $p$ and $q$, or 
alternatively $p^2$ and ${p'}^2 = (p+q)^2$). This means Borel transformation 
should give us an expression for the three-point function in terms of two Borel
masses, $M^2$ and ${M'}^2$. If the baryon masses were close, we could use $M^2 
= {M'}^2$. But in this case, we are behooved to consider different values of 
the Borel masses. Following \cite{Nielson2000}, we assume that they should obey
a ratio
\begin{eqnarray}
\label{BorelRatio}
\frac{{M'}^2}{{M}^2} = \frac{M_{B'}^2}{M_{B}^2},
\end{eqnarray}
where $M_B$ and $M_{B'}$ are respective Lambda baryon ($\Lambda_c$ and 
$\Lambda_s$) masses in this case. This helps us to express the $\Pi_{3c}(p,q)$ 
in Eq.~(\ref{Pi3cEvaltd_ZeroOne_Final_redefined}) in terms of one variable 
$p^2$. We define a quantity $\delta$ as
\begin{eqnarray}
\label{def_delta}
\delta \equiv \left( \frac{M_{B'}^2}{M_B^2} - 1 \right)
\end{eqnarray}
to write, in the limit of zero pion mass,
\begin{eqnarray}
\label{exp1}
q.(p+\gamma q) &=& q.p = \frac{1}{2}(p'^2 - p^2) = \frac{1}{2} \delta p^2, \\
\label{exp2}
(p + \gamma q)^2 &=& (1 + \gamma \delta)p^2, \\
\label{exp3}
(1 - \gamma)p^2 + \gamma {p'}^2 &=& (1 + \gamma \delta)p^2, \\
\label{exp4}
p.(p+ \gamma q) &=& (1 + \frac{\gamma \delta}{2})p^2,
\label{exp5}
\end{eqnarray} 
Also, we can express $a$ and $a'$ in a form that is convenient for Borel 
transform:
\begin{eqnarray}
a(\rho, \kappa, \gamma)
&=&
c_1(\rho, \kappa, \gamma)
\Big[
p^2 - b(\rho, \kappa, \gamma)^2
\Big] ~,\\
a'(\rho, \gamma)
&=&
c_2(\rho, \gamma)
\Big[
p^2 - {b'(\rho, \gamma)}^2
\Big]
\end{eqnarray}
where
\begin{eqnarray}
c_1(\rho, \kappa, \gamma)
&=&
\Big\{
\frac{\kappa \rho (1 - \rho)}{g(\rho, \kappa)}
\Big\}
\left[
(1 - \kappa) + \frac{1}{2} \gamma \delta g'(\rho, \kappa)
\right],\\
b^2(\rho, \kappa, \gamma)
&=&
\frac{g(\rho, \kappa){\mathcal{Z}}^2 }{(1 - \rho) \left[ (1 - \kappa) + 
\frac{1}{2} \gamma \delta g'(\rho, \kappa) \right]}, \\
c_2(\rho, \gamma) &=& \Big[- \frac{\delta \gamma \rho (1 - \rho)}{2} \Big], \\
{{b'}^2(\rho, \gamma)}
&=& -\frac{2}{\gamma \delta (1 - \rho)} {\mathcal{Z}}^2, \\
g'(\rho, \kappa) &=& (1 - \kappa) - \kappa \rho (1 - \rho)~.
\end{eqnarray}
Using $P^2 = - p^2$ and applying the Borel transformation
\begin{eqnarray}
\label{BT}
{\mathcal{B}}_{M^2}  = \lim_{P^2, n \rightarrow \infty; P^2/n = M^2} 
\frac{(P^2)^{n+1}}{n!} \left( - \frac{d}{dP^2} \right)^n,
\end{eqnarray}
we find
\begin{eqnarray}
{\mathcal{B}}_{M^2} \left[  \ln (P^2 + b^2) \right] &=& - M^2 e^{- b^2/M^2}, \\
{\mathcal{B}}_{M^2} \left[ P^2 \ln (P^2 + b^2) \right] &=&  M^2 (b^2 + M^2) 
e^{- b^2/M^2}, \\
{\mathcal{B}}_{M^2} \left[ (P^2)^2 \ln (P^2 + b^2) \right] &=& - M^2 (2M^4 + 2 b^2
 M^2 + b^4) e^{- b^2/M^2}
\end{eqnarray}
to write the Borel-transformed function $\tilde{\Pi}_{3c}(M)=
{\mathcal{B}}_{M}\left[\Pi_{3c}(P; P^2 = -p^2) \right]$ as
\begin{eqnarray}
\label{Pi3COPEFinal}
{\tilde{\Pi}}_{3Q}(M)
&=&
\frac{1}{2(4 \pi)^4} \int_0^1 d \gamma d \rho d \kappa \frac{\delta \rho^3 
(1 - \rho)^2}{g^3} \Big[2\Big( \kappa(1 + \gamma \delta) - c_1 \rho \Big)M^6 
 \nonumber\\
&& \qquad \qquad
+ \Big(2 \kappa(1 + \gamma \delta) - c_1 \rho  \Big) M^4 b^2
+ \kappa (1 + \gamma \delta) M^2 b^4
\Big]e^{- b^2/M^2} \nonumber\\
&&
+\frac{1}{2(4 \pi)^4} \int_0^1 d \gamma d \rho~ \Big(\frac{\delta}{1-\rho}\Big)
 \left[ -2 c_2 M^6 - c_2 {b'}^2 M^4 \right] e^{-{b'}^2/M^2}~.
\end{eqnarray}
What we have got finally in Eq.~(\ref{Pi3COPEFinal}) is the Operator Product 
Expansion (OPE) of the three-point correlator. This expression is evaluated 
using Mathematica. We will equate this expression to a phenomenological model 
for the decay process in order to calculate the coupling, $g_{\Lambda_c 
\rightarrow \Lambda_s \pi}$~.

\section{Phenomenological side of the decay}
We obtain the phenomenological side for this process from the restrictions 
imposed by symmetry. To illustrate that, let's recall the three-point function 
here:
\begin{eqnarray}
\label{def_pi3}
\Pi_3(p,q) &=&
i \int d^4 x d^4 y~ e^{ip.x + iq.y}~ \langle 0 \vert T \Big[
\eta_{\Lambda_c}(x) H_W(y) \eta_{\Lambda_s}(0)
\Big] \vert 0 \rangle~.
\end{eqnarray}
We can express this function in terms of physical intermediate states of our 
interest through the following matrix elements:
\begin{eqnarray}
\label{def_lambda_LambdaC}
\langle 0 \vert \eta_{\Lambda_c} \vert \Lambda_c(p) \rangle &=& \lambda_{\Lambda_c} 
u(p), \\
\label{def_lambda_LambdaS}
\langle \Lambda_s(p') \vert \eta_{\Lambda_s} \vert 0 \rangle &=& \lambda_{\Lambda_s}
 \bar{u}(p'), \\
\label{def_g}
\langle \Lambda_c(p) \vert j^\mu \vert \Lambda_s(p') \rangle &=&
g(p,p')  \left[ \bar{u}(p) i \gamma^\mu  u(p') \right],
\end{eqnarray}
where $\lambda_{\Lambda_c}$ and $\lambda_{\Lambda_s}$ are the couplings of the 
charmed and the strange lambda baryon currents to their hadronic states $u(p)$ 
is a spinor obeying the normalization $u(p) \bar{u}(p) = 2 m_B$ with $m_B$ 
being the mass of the the baryon $B$, and $g(p,p')$ is the coupling of the pion
 current to the baryons and is related to the dimensionless coupling constant, 
$g_{\Lambda_c \rightarrow \Lambda_s \pi}$ that we seek to find out in this paper, 
through the following relation\cite{RRY85}:
\begin{eqnarray}
\label{def_g_dimless}
g(p,p') = g_{\Lambda_c \rightarrow \Lambda_s \pi} \left[ \frac{2 m_\pi^2 f_\pi}{\left(m_u 
+ m_d \right) \left( q^2 - m_\pi^2 \right)} \right],
\end{eqnarray}
where $m_\pi$, $m_u$ and $m_d$ are the mass of pion, up and down quarks, $f_\pi$ 
is the pion decay constant, and $q^2 = (p'-p)^2$. Using Eq.
~(\ref{def_lambda_LambdaC}), (\ref{def_lambda_LambdaS}), (\ref{def_g}), 
(\ref{def_g_dimless}) and (\ref{HW}), we get
\begin{eqnarray}
\label{Pi3Pheno1}
\Pi_3^{\rm{pheno}}(p,p')
&=&
iG_F F_\pi V_{ud} 
~
\left[
\frac{\langle 0 \vert \eta_{\Lambda_c} \vert \Lambda_c(p) \rangle }
{\not{p} - m_{\Lambda_c}} \right] 
\langle \Lambda_c(p) \vert j^\mu \vert \Lambda_s(p') \rangle
\left[
\frac{
\langle \Lambda_s(p') \vert \eta_{\Lambda_s} \vert 0 \rangle}
{\not{p'} - m_{\Lambda_s}} 
\right] \nonumber\\ 
&=&
4i G_F ~g_{\Lambda_c \rightarrow \Lambda_s \pi} ~V_{ud}
\frac{2 \lambda_{\Lambda_c} \lambda_{\Lambda_s} m_{\Lambda_c} m_{\Lambda_s} m_\pi^2 F_\pi^2 }
{\left(m_u+ m_d \right)  \left( q^2 - m_\pi^2 \right)} ~
\frac{\left( \not{p}+m_{\Lambda_c} \right) \not{q} \left(\not{p'} + m_{\Lambda_s} 
\right)}{(p^2 - m_{\Lambda_c}^2)({p'}^2 - m_{\Lambda_s}^2)}
\end{eqnarray}
But,
\begin{eqnarray}
\label{DiffLorentzStruc}
(\not{p} + m_{\Lambda_c}) \not{q} (\not{p'} + M_{\Lambda_s})
&=&
\left( m_{\Lambda_c} + m_{\Lambda_s} \right) p.q +
(m_{\Lambda_c} m_{\Lambda_s} - p^2)\not{q} + 2q.p \not{p}  \nonumber\\
&& \qquad -i \left[
q_\mu \sigma^{\mu \nu} p_\nu m_{\Lambda_c} + p_\mu \sigma^{\mu \nu} q_\nu m_{\Lambda_s}
\right]
\end{eqnarray}
Out of all the terms present in Eq.~(\ref{DiffLorentzStruc}), we can only 
concentrate on the first one and ignore the others because only the first term 
is consistent with the Lorentz structure of the OPE side of the three-point 
function. Inserting it back to Eq.~(\ref{Pi3Pheno1}), and considering 
$m_u \approx m_d = m_q$, we get
\begin{eqnarray}
\Pi_3^{\rm{pheno}}(p,p')&=&
- \frac{i~ g_{\Lambda_c \rightarrow \Lambda_s \pi}~ G_F V_{ud} \lambda_{\Lambda_c} 
\lambda_{\Lambda_s} m_{\Lambda_c} m_{\Lambda_s} F_\pi^2 }{ m_q } ~
\left[\frac{\delta p^2 \left( m_{\Lambda_c} + m_{\Lambda_s} \right)}{(p^2 - 
m_{\Lambda_c}^2)({p'}^2 - m_{\Lambda_s}^2)} \right].
\end{eqnarray}
In the above expression, we have considered $q^2 = 0$ and $\delta$ is defined 
in Eq.~(\ref{def_delta}). Using Eq.~(\ref{BorelRatio}), we get
\begin{eqnarray}
\Pi_3^{\rm{pheno}}(p) &=&  - \frac{i g_{\Lambda_c \rightarrow \Lambda_s \pi}~ G_F V_{ud} 
\lambda_{\Lambda_c} \lambda_{\Lambda_s} m_{\Lambda_c} m_{\Lambda_s} F_\pi^2 }
{ m_q }\nonumber\\&&
~ \times \frac{\delta \left( m_{\Lambda_c} + m_{\Lambda_s} \right) m_{\Lambda_c}^2}
{m_{\Lambda_s}^2} \left[\frac{1}{p^2 - m_{\Lambda_c}^2} + \frac{m_{\Lambda_c}^2}
{(p^2 - m_{\Lambda_c}^2)^2}\right] ~.
\end{eqnarray}
Now, we apply the Borel transformation [Eq.~(\ref{BT})] on $\Pi_3^{\rm{pheno}}(p)$
to get
\begin{eqnarray}
\label{Pi3CPhenoFinal}
{\mathcal{B}}_{M^2} \left[
\Pi_3^{\rm{pheno}}(p) 
\right]
&=&i g_{\Lambda_c \rightarrow \Lambda_s \pi} G_F V_{ud} \lambda_{\Lambda_c} \lambda_{\Lambda_s}
\frac{~\delta  \left( m_{\Lambda_c} + m_{\Lambda_s} \right)  F_\pi^2 }
{ m_q}~ \frac{m_{\Lambda_c}^2}{ m_{\Lambda_s}^2}
\left(
1 - \frac{m_{\Lambda_c}^2}{M^2}
\right) e^{- \frac{m_{\Lambda_c}^2}{M^2}}, \nonumber\\
\end{eqnarray}
where we consider $m_{\Lambda_s} = 1.115 ~\rm{GeV}$, $m_{\Lambda_c} =
2.286~\rm{GeV}$, $m_s = 0.095~\rm{GeV}$, $m_c = 1.275~\rm{GeV}$, $F_\pi = 
0.092~\rm{GeV}$, $m_q = 0.004~\rm{GeV}$. For the couplings, $\lambda_{\Lambda_c}$ 
and $\lambda_{\Lambda_s}$, we follow \cite{Nielson1999}, where the values of 
these parameters were obtained from baryonic mass sum rules in heavy quark 
effective theory \cite{NarisonEtAl1992}:
\begin{eqnarray}
\label{couplingB}
2 (4 \pi)^4 ~\vert \lambda_{B} \vert^2 e^{- M_{B}^2/M^2}
&=&
M^6 E_2^B + \frac{2}{3} a m_Q (1 - 3 \gamma) M^2 E_0^B + b M^2 E_0^B + 
\frac{4}{9} a^2 (3 + 4 \gamma), \nonumber\\
\end{eqnarray}
where $B$ denotes the baryon $\Lambda_c$ or $\Lambda_s$, $m_Q$ denotes the mass
of the heavy quark (charm or strange), and
\begin{eqnarray}
a &=& - (2\pi)^2 \langle \bar{q} q \rangle \approx 0.5 ~\rm{GeV}^3, \\
b &=& \pi^2 \langle \alpha_s G^2/\pi \rangle \approx 0.12 ~\rm{GeV}^4, \\
\gamma &=& \langle \bar{q} q \rangle /\langle \bar{s} s \rangle - 1 \approx -0.2
\end{eqnarray}
and $E_n^B$ represents the continuum contribution,
\begin{eqnarray}
E_n^B = 1 - \left( 1 + x + \frac{x^2}{2} + \cdots + \frac{x^n}{n!} \right) e^{-x}
\end{eqnarray}
with $x = s_{B}/M^2$, $s_B$ being the continuum threshold. Notice that, we can 
only determine the absolute value of $g_{\Lambda_c \rightarrow \Lambda_s \pi}$ and not 
the sign, since Eq.~(\ref{couplingB}) gives only the absolute value of 
$\lambda_B$. Note that the the couplings $\lambda_{\Lambda_c}$ and
$\lambda_{\Lambda_s}$ are so well known that they do not add to the errors
estimated from the Borel Mass plots.

\section{Results}
In this section, we compare the OPE side of the three-point function given in 
Eq.~(\ref{Pi3COPEFinal}) to the phenomenological side in 
Eq.~(\ref{Pi3CPhenoFinal}) to find out the coupling. The free parameters 
appearing in the latter expression are the continuum thresholds, $s_{B}$ and 
$s_{B'}$, that determine the couplings, $\lambda_B$ and $\lambda_{B'}$. From 
Eq.~(\ref{couplingB}), we can write
\begin{eqnarray}
\label{Compare1}
\vert \lambda_B/\lambda_{B'} \vert ^2~ e^{-(M_B^2 - M_{B'}^2)/M^2}
&=&
\frac{E_2^{B} + \left(\frac{2 a m_c}{3 M^4} \right) (1 - 3 \gamma) E_0^B + 
\left(\frac{b}{M^4}\right) E_0^B + \left(\frac{4a^2}{9M^6}\right)(3 + 4 \gamma)}
{E_2^{B'} + \left(\frac{2 a m_{s}}{3 M^4} \right) (1 - 3 \gamma) E_0^{B'} + 
\left(\frac{b}{M^4}\right) E_0^{B'} + \left(\frac{4a^2}{9M^6}\right)(3 + 4 
\gamma)}~.\nonumber\\
\end{eqnarray}
For what follows from here, $B$ denotes $\Lambda_c$, $B'$ denotes $\Lambda_s$. 
To estimate $s_B$, $s_{B'}$, $\lambda_B$ and $\lambda_{B'}$, we start with a 
residual
\begin{eqnarray}
\label{DefRes}
R(M^2) = (\rm{rhs} - \rm{lhs})^2/\rm{lhs}^2,
\end{eqnarray}
where rhs and lhs denotes the right- and left-hand sides of 
Eq.~(\ref{Compare1}) respectively. Then, we assume a prior range of values for 
the free parameters and minimize the residual in Eq.~(\ref{DefRes}). We define 
\begin{eqnarray}
r &=& \vert \lambda_B/\lambda_{B'} \vert ^2, \\
s_B &=& \left( M_B + \Delta \right)^2, \\
s_{B'} &=& \left( M_{B'} + \Delta \right)^2~.
\end{eqnarray}
The above definitions of $s_B$ and $s_{B'}$ help us to constrain the continuum 
threshold with just one free parameter, $\Delta$. Similar parametrization has 
been adopted in other works, $e.g.$, in \cite{Nielson1999} and 
\cite{Nielson2000}. Now, using this residual method, we attempt to find a Borel
 window over which (i) the residual will be close to zero, (ii) it will 
effectively become constant over this range (window) of $M$, because our result
 should be independent of the extra parameter $M$ that we introduced just to 
regulate the divergences in the three-point function. We start with a prior 
range of the free parameters: $\Delta \in \left[ 0.5~{\rm{GeV}}, 0.9~{\rm{GeV}}
 \right]$, $r \in \left[1, 18 \right]$ and attain a suitable Borel window for 
the following values of the free parameters:
\begin{eqnarray}
\label{EstimaterAndDelta}
r &=& 17.259,~\Delta = 0.784~\rm{GeV}~.
\end{eqnarray}

\begin{figure}
\begin{center}
\epsfig{file=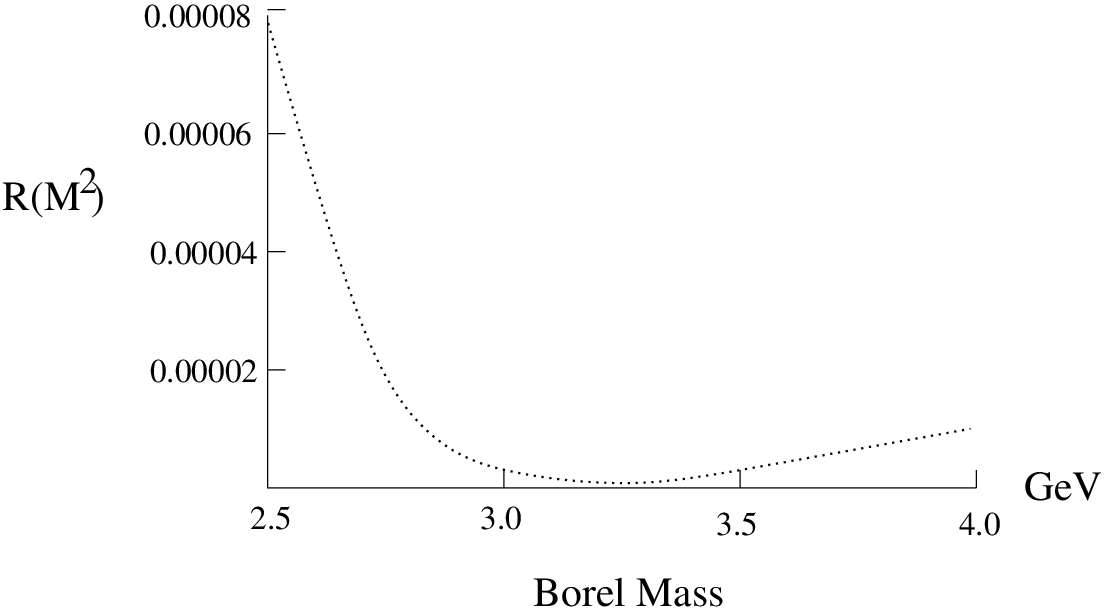,height= 6cm,width=10cm}
\end{center}
%\label{Figure 1}
\caption{Residual of Eq. 51 as a function of Borel mass for $r = 17.259$, 
$\Delta = 0.784~\rm{GeV}$}
\end{figure}

The residual is plotted in Figure 2 for the above-mentioned values of the free 
parameters. Using the values of $\Delta$ obtained in 
Eq.~(\ref{EstimaterAndDelta}) in Eq.~(\ref{Pi3CPhenoFinal}), and comparing 
Eq.~(\ref{Pi3CPhenoFinal}) with Eq.~(\ref{Pi3COPEFinal}), we get $g_{\Lambda_c 
\rightarrow \Lambda_s \pi}$ as a function of Borel mass, as shown in Figure 3. 
Figure 2 and 3 allows us to choose a Borel window $M \in \left[ 2.7~{\rm{GeV}},
 3.2~{\rm{GeV}} \right]$ over which we estimate the value of the coupling 
constant, $g_{\Lambda_c \rightarrow \Lambda_s \pi}$.
\begin{eqnarray}
g_{\Lambda_c \rightarrow \Lambda_s \pi} 
&=&
1.060 \pm 0.014
\end{eqnarray}

\begin{figure}
\begin{center}
\epsfig{file=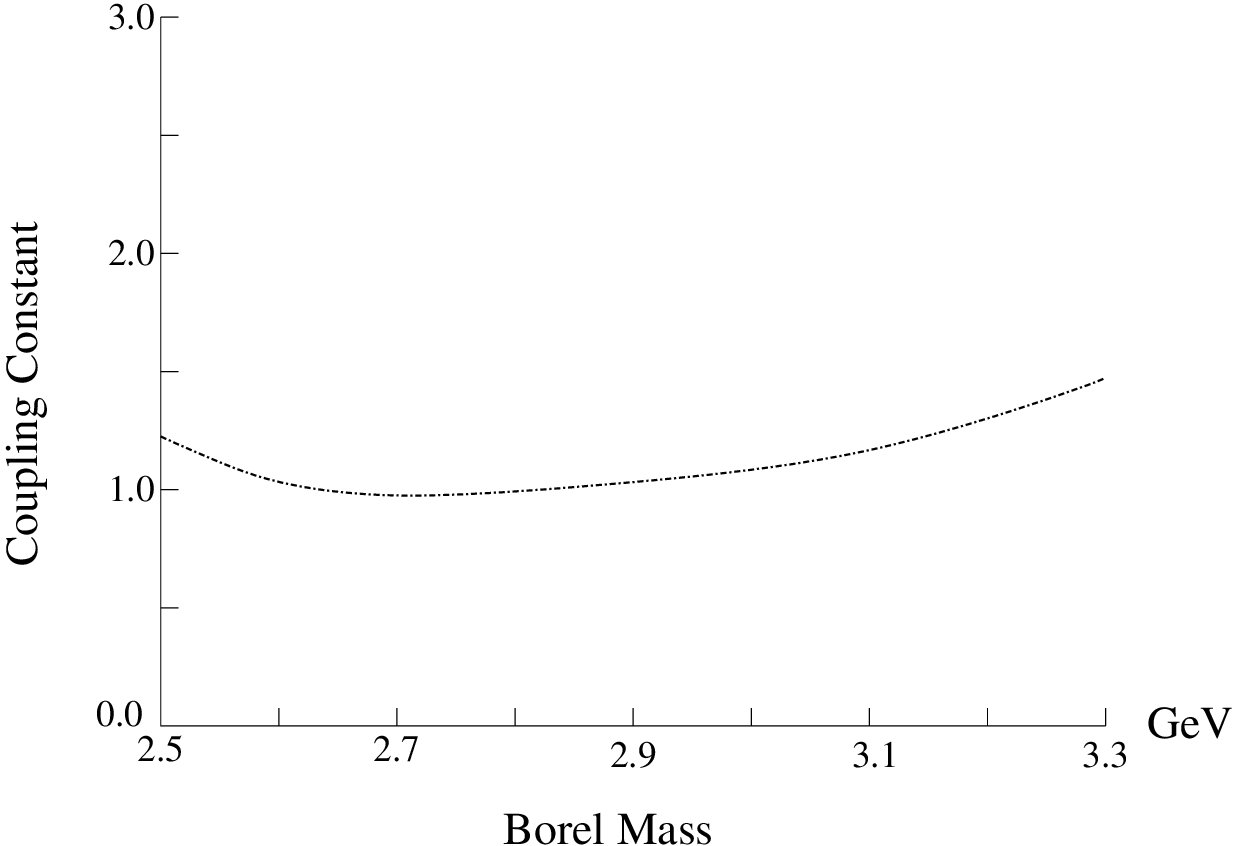,height= 6cm,width=10cm}
\end{center}
\label{Figure 1}
\caption{Coupling constant $g_{\Lambda_c \rightarrow \Lambda_s \pi}$ as a function of 
Borel mass}
\end{figure}

\section{Conclusion}
In this work, we have calculated the coupling constant, $g_{\Lambda_c 
\rightarrow \Lambda_s \pi}$, for the decay process 
$\Lambda_c^+ \rightarrow \Lambda^0 \pi^+$. We used a parametric representation 
of the propagators analytically continued to complex D-dimensions in order to 
solve the lowest order perturbative diagram corresponding to this process. We 
achieved convergence with a reduced Borel transformation. This method will be 
extended to estimate other Cabibbo-favored and Cabibbo-suppressed decays of 
heavy-quark baryons, and also to estimate decays for other weak modes of 
charmed lambda in future.

\vspace{1cm} 
\Large{{\bf Acknowledgements}}\\
\normalsize
This work was supported in part by a grant from the Pittsburgh Foundation
and in part by the Carnegie Mellon University Department of Physics.
\vspace{2mm}

\begin{appendix}
\setcounter{equation}{0}
\renewcommand{\theequation}{A.\arabic{equation}}

\Large
{\bf Appendix A: OPE side of the three-point function}
\normalsize
\vspace{2mm}

Using Eq.~(\ref{lambdacurrent}), (\ref{HW}) and (\ref{quark-J}) and keeping the
 essential terms, $\Pi_3(x,y)$ is
\beq
\label{A.1}
 \Pi_3(x,y)&=& iG_F F_\pi V_{ud}~ q_\alpha \langle 0 \vert ~T 
\Big[\epsilon^{abc}\bar{u}(x)^a 
C\gamma_\mu d^b(x) \gamma_5 \gamma^\mu c(x)^c s^j(y) \gamma^\alpha(1-\gamma_5) 
\bar{c}^j(y) V_{cs} \nonumber \\
  && \qquad \qquad \epsilon^{def} u^d(0)C  \gamma_\lambda \bar{d}^e(0) 
\gamma_5 \gamma^\lambda \bar{s}^f (0) \Big]~\vert 0 \rangle \; .
\eeq

\noindent
From Eq.~(\ref{3pt-correlator}), (\ref{A.i}), and using $q(x)=\int 
\frac{d^4 k}{(2\pi)^4} e^{-ik \cdot x}q(k)$,
\beq
\label{3pt-correlator-p-q}
\Pi_3(p,q) &=& iG_F F_\pi V_{cs}V_{ud} ~q_\alpha \int d^4x d^4y 
e^{ip\cdot x} e^{iq \cdot y} \int \frac{d^4 k_1}{(2\pi)^4}\frac{d^4 k_2}
{(2\pi)^4} \frac{d^4 k_3}{(2\pi)^4} \frac{d^4 k_4}{(2\pi)^4}  \\
 &&e^{ik_1 \cdot x}e^{-ik_2 \cdot x}e^{-ik_3 \cdot (x-y)}e^{-ik_4 \cdot y}
Tr[S_u(k_1) C\gamma_\mu S_d(k_2) C^* \gamma_\lambda \gamma_5 \gamma^\mu  
S_c(k_3) \gamma^\alpha (1-\gamma_5) S_s(k_4)\gamma^\lambda \gamma^5] \nonumber
 \; ,
\eeq 
where the quark propagator is $S_q(k)=(\not k+m_q)/(k^2-m_q^2)=
 (k_\mu \gamma^\mu+m_q)/(k^2-m_q^2)$.

\noindent
Using $m_u,m_d \ll m_c$, the trace in Eq(\ref{3pt-correlator-p-q}) is
\beq
\label{trace}
Tr[S_u(k_1)C\gamma_\mu S_d(k_2)\gamma^\nu C\gamma_5 \gamma^\mu S_c(k_3)
S_s(k_4)\gamma_\nu \gamma_5]&=& Tr[\not k_1 C\gamma_\mu \not k_2 C^* 
\gamma_\lambda \gamma_5 \gamma^\mu (\not k_3 +m_c) \gamma^\alpha (1-\gamma_5)
 \nonumber\\
&&(\not k_4 +m_s) \gamma_\lambda \gamma^5]\frac{1}{k_1^2 k_2^2 (k_3^2-m_c^2)
(k_4^2-m_s^2)}
\eeq

\noindent
In carrying out the trace in Eq(\ref{trace}) note that $Tr[\gamma_5 \gamma_\nu
\gamma_\lambda] = 0$, and one obtains for the trace on the right hand side
\beq
\label{trace2} 
TR &=& Tr[\not k_1 C\gamma_\mu 
\not k_2 C^* \gamma_\lambda \gamma_5 \gamma^\mu (\not k_3 +m_c) \gamma^\alpha 
(1-\gamma_5) (\not k_4 +m_s) \gamma_\lambda \gamma_5] \nonumber\\
&=& 16 m_c (k_1\cdot k_2 k_4^\alpha +k_2\cdot k_4 k_1^\alpha -k_1\cdot k_4 
k_2^\alpha) \nonumber \\
&&\qquad \qquad 
-16 m_s (-k_1\cdot k_2 k_3^\alpha +
k_2\cdot k_3 k_1^\alpha -k_1\cdot k_3 k_2^\alpha) \; .
\eeq

\noindent
Making use of $\int d^4x e^{ix \cdot (p-k)}=(2\pi)^4 \delta^{(4)}(p-k)$ so $k_3 = p + k_1 - k_2$
and $k_4 = p + k_1 - k_2 + q$, one obtains
\beq
\label{3pt-correlator-p-final}
\Pi_3(p,q) &=& 16 i G_F F_\pi V_{cs} V_{ud}
\int \frac{d^4 k_1}{(2\pi)^4}\frac{d^4 k_2}{(2\pi)^4} (m_c F_1 + m_s F_2) 
\nonumber \\
&& \times \frac{1}{[k_1^2 k_2^2((p +k_1 - k_2)^2-m_c^2)((p + q + k_1 - k_2)^2-m_s^2)]} \; ,
\eeq
where
\beq
\label{F1}
F_1 &=&  k_1.q ~k_2.(p+q+k_1-k_2) - k_2.q~k_1.(p+q+k_1-k_2) + k_1.k_2~q.(p+q+k_1-k_2),\qquad \\
F_2 &=& k_1.(p+k_1 - k_2) ~k_2.q - k_2.(p+k_1-k_2)~k_1.q + k_1.k_2~q.(p+k_1-k_2)
\eeq
Using $(k_1 + p) \equiv l$ and $k_2 \equiv k$, $\Pi_3(p,q)$ in 
Eq(\ref{3pt-correlator-p-final}) can be expressed as
\beq
\label{Pi3in3cAnd3s}
\Pi_{3}(p, q) &=&
16 i G_F F_\pi V_{ud} V_{cs} ~ \big(m_c \Pi_{3c} +  m_s \Pi_{3s} 
\big) 
\; ,
\eeq
with 
\beq
\label{Pi3c1}
\Pi_{3c}(p,q) &=& \int \frac{d^4 k}{(2 \pi)^4} \frac{1}{k^2}
\int \frac{d^4 l}{(2 \pi)^4} \frac{F_1}{(l+p)^2 \left[ (l-k)^2 - m_c^2  \right]
\left[ (l- k - q)^2 - m_s^2 \right]} 
\eeq 
\beq
\label{Pi3s1}
\Pi_{3s}(p,q) &=& \int \frac{d^4 k}{(2 \pi)^4} \frac{1}{k^2} 
\int \frac{d^4 l}{(2 \pi)^4}\frac{F_2}{(l+p)^2 \left[ (l-k)^2 - m_c^2  \right] 
\left[ (l - k - q)^2 - m_s^2 \right]}
\; ,
\eeq
where
\beq
\label{F1z}
F_1 =  k\cdot (l-p) q\cdot(l-k+q) +k\cdot(l-k+q)q\cdot(l-p) -
(l-p)\cdot (l-k+q) q\cdot k  
\eeq
\beq
\label{F2z}
F_2 =  k\cdot (l-p)~ q\cdot(l-k) - k\cdot(l-k)~q\cdot(l-p) +
(l-p)\cdot (l-k) ~q\cdot k \; .
\eeq
Here we consider $m_s \ll m_c$ to write
\begin{eqnarray}
\Pi_3(p,q) &=& 
16 i m_c G_F F_\pi V_{ud} V_{cs}~ \Pi_{3c} (p,q)
\end{eqnarray}

Defining
\begin{eqnarray}
\label{PimunuomegaAppendix}
\tilde{\Pi}^{\mu \nu \omega}(p,q)
&=&
\int \frac{d^4 k}{(2\pi)^4} \frac{k_\mu}{k^2}
\int \frac{d^4 l}{(2 \pi)^4}
\frac{(l-p)^\nu (l - k + q)^\omega}{(l+p)^2 \left[(l - k)^2 - m_c^2  \right] \left[  (l - k + q)^2 - m_s^2\right]} \nonumber\\
&=&
\int \frac{d^4 k}{(2\pi)^4} \frac{k_\mu}{k^2}
\int \frac{d^4 l}{(2 \pi)^4}
\frac{l^\nu (l - k + p + q)^\omega}{l^2 \left[(l - k + p)^2 - m_c^2  \right] \left[  (l - k + p + q)^2 - m_s^2\right]}
\end{eqnarray}
to write
\begin{eqnarray}
\label{Pi3c}
\Pi_{3c}(p,q)
&=&
q_\nu \left[
\tilde{\Pi}^{\mu \mu \nu} + \tilde{\Pi}^{\mu \nu \mu} - \tilde{\Pi}^{\nu \mu \mu}
\right]
\end{eqnarray}

Using $k \rightarrow - k$, $l \rightarrow - l$, we get
\begin{eqnarray}
\Pi^{\mu \nu \omega}(p,q) = - \tilde{\Pi}^{\mu \nu \omega}
&=&
\int \frac{d^4 k}{(2\pi)^4} \frac{k_\mu}{k^2}
\int \frac{d^4 l}{(2 \pi)^4}
\frac{l^\nu (l - k -p - q)^\omega}{l^2 \left[(l - k - p)^2 - m_c^2  \right] \left[  (l - k - p - q)^2 - m_s^2\right]}. \qquad
\end{eqnarray}
Thus, 
\begin{eqnarray}
\Pi_3(p,q) &=& 
-16 i m_c G_F F_\pi V_{ud} V_{cs}~ \Pi_{3Q} (p,q)
\end{eqnarray}
where
\begin{eqnarray}
\Pi_{3Q}(p,q)
&=&
q_\nu \left[
\Pi^{\mu \mu \nu} + \Pi^{\mu \nu \mu} - \Pi^{\nu \mu \mu}
\right]
\end{eqnarray}
\end{appendix}
\vspace{2mm}

\begin{appendix}
\setcounter{equation}{0}
\renewcommand{\theequation}{B.\arabic{equation}}

\LARGE
{\bf Appendix B: Evaluating $\Pi^{\mu \nu \omega}(p,q)$}
\normalsize
\vspace{2mm}

We define 
\begin{eqnarray}
\Pi_{l0}(p,q)
&=&
\int \frac{d^4 l}{(2 \pi)^4}~
\frac{1}{l^2 \left[(l - k - p)^2 - m_c^2  \right] \left[  (l - k - p - q)^2 - m_s^2\right]} \nonumber\\
\Pi_{l1}^{\mu}(p,q)
&=&
\int \frac{d^4 l}{(2 \pi)^4}~
\frac{l^\mu}{l^2 \left[(l - k - p)^2 - m_c^2  \right] \left[  (l - k - p - q)^2 - m_s^2\right]} \nonumber\\
\Pi_{l2}^{\mu \nu}(p,q)
&=&
\int \frac{d^4 l}{(2 \pi)^4}~
\frac{l^\mu l^\nu}{l^2 \left[(l - k - p)^2 - m_c^2  \right] \left[  (l - k - p - q)^2 - m_s^2\right]}
\end{eqnarray}
These integrals are evaluated using the regularization technique adopted in \cite{kb17} to give us
\begin{eqnarray}
\Pi_{l0}(p,q)
&=&
\frac{1}{(4 \pi)^2} \int_0^\infty d \alpha d \beta d \gamma ~
\frac{e^{A(\alpha, \beta, \gamma)+ \beta m_c^2 + \gamma m_s^2}}{(\alpha+\beta+\gamma)^2}, \\
\Pi_{l1}^{\mu}(p,q)
&=&
\frac{1}{(4 \pi)^2} \int_0^\infty d \alpha d \beta d \gamma ~
\frac{[(\beta+\gamma)k + s]^\mu}{(\alpha+\beta+\gamma)^3}~
e^{A(\alpha, \beta, \gamma)+ \beta m_c^2 + \gamma m_s^2}, \\
\Pi_{l2}^{\mu \nu}(p,q)
&=&
\frac{1}{(4 \pi)^2} \int_0^\infty d \alpha d \beta d \gamma ~
e^{A(\alpha, \beta, \gamma)+ \beta m_c^2 + \gamma m_s^2} \Big[
- \frac{1}{2} \frac{g^{\mu \nu}}{(\alpha+\beta+\gamma)^3} \nonumber\\&& \qquad
+ \frac{[(\beta+\gamma)k + s]^\mu~[(\beta+\gamma)k + s]^\nu}{(\alpha+\beta
+\gamma)^4} \Big] {\rm \;where\;}
\end{eqnarray}

\begin{eqnarray}
s^\mu &=& (\beta+\gamma)p^\mu + \gamma q^\nu~ \\
A(\alpha, \beta, \gamma)&=& (p+k)^2 (\frac{\beta^2}{\alpha+\beta+\gamma}-\beta)
+(p+k+q)^2(\frac{\gamma^2}{\alpha+\beta+\gamma}-\gamma) \nonumber \\
&& +\frac{2 \beta \gamma (p+k)\cdot(p+k+q)}{\alpha+\beta+\gamma} \; .
\end{eqnarray}

Using $d^4 k \rightarrow d^D k$, $D \equiv 4 - 2 \epsilon$, $\beta \rightarrow \rho \beta$, $\gamma \rightarrow \rho \gamma$, $\delta(\rho - \beta - \gamma) \rightarrow \delta (1 -\beta - \gamma )/\rho$, and repeating for $\alpha$ as well, one obtains
\begin{eqnarray}
\label{Pil0}
\Pi_{l0}(p,q)
&=&
\frac{1}{(4 \pi)^2} \int_0^1 d \gamma d \rho \int_0^\infty d \kappa~
\rho e^{A+ \kappa \rho[(1 - \gamma)m_c^2 + \gamma m_s^2]}, \\
\label{Pil1mu}
\Pi_{l1}^{\mu}(p,q)
&=&
\frac{1}{(4 \pi)^2} \int_0^1 d \gamma d \rho \int_0^\infty d \kappa~
\rho^2 [k+p+\gamma q]^\mu~ 
e^{A+ \kappa \rho[(1 - \gamma)m_c^2 + \gamma m_s^2]},\\
\label{Pil2munu}
\Pi_{l2}^{\mu \nu}(p,q)
&=&
\frac{1}{(4 \pi)^2} \int_0^1 d \gamma d \rho \int_0^\infty d \kappa~
~\rho 
\left[
- \frac{1}{2 \kappa}g^{\mu \nu} + \rho^2 (k+p+\gamma q)^\mu (k+p+\gamma q)^\nu
\right]
e^{A+ \kappa \rho[(1 - \gamma)m_c^2 + \gamma m_s^2]} \nonumber\\
\end{eqnarray}
where we redefine
\begin{eqnarray}
A &=&
D k^2 + F.k + f(p,q)
\end{eqnarray}
where
\begin{eqnarray}
D &=& - \kappa \rho (1 - \rho),\\
F^\mu &=& -2 \kappa \rho (1 - \rho) (p+\gamma q)^\mu, \\
f(p,q) &=& - \kappa \rho (1 - \rho) p.(p+ \gamma q)
\end{eqnarray}
Using Eq.~(\ref{Pil0}), (\ref{Pil1mu}), (\ref{Pil2munu}) in Eq.~(\ref{Pimunuomega}), we get
\begin{eqnarray}
\label{Pimumunu}
\Pi^{\mu \mu \nu}(p,q)
&=&
\int \frac{d^4 k}{(2 \pi)^4}~ \frac{k_\mu}{k^2} 
\int \frac{d^4 l}{(2 \pi)^4}~
\frac{l^\mu (l - k - p - q)^\nu}{l^2 \left[ (l-k-p)^2 - m_c^2 \right] \left[ (l - k - p - q)^2 - m_s^2  \right]} \nonumber\\
&=&
\int \frac{d^4 k}{(2 \pi)^4}~ \frac{k_\mu}{k^2} ~
\Big[
\Pi_{l2}^{\mu \nu} - \left( k+p+q \right)^\nu \Pi_{l1}^\mu
\Big] \nonumber\\
&=&
\frac{1}{(4 \pi)^2} \int_0^\infty d \alpha d \beta d \gamma~ e^{A+\beta m_c^2 + \gamma m_s^2}
\int \frac{d^4 k}{(2 \pi)^4}~ \frac{k_\mu}{k^2} ~
\Big[
- \frac{g^{\mu \nu}}{2(\alpha+\beta+\gamma)^3} \nonumber\\ && \qquad
+ \frac{\left[ (\beta+\gamma)k + s \right]^\mu \left[ (\beta+\gamma)k+s \right]^\nu}{(\alpha+\beta+\gamma)^4}
- \frac{(k+p+q)^\nu [(\beta+\gamma)k+s]^\mu}{(\alpha+\beta+\gamma)^3}
\Big] \nonumber\\
\end{eqnarray}
Similarly, we can write
\begin{eqnarray}
\label{Pimunumu}
\Pi^{\mu \nu \mu}
&=&
\int \frac{d^4 k}{(2 \pi)^4}~ \frac{k_\mu}{k^2} 
\int \frac{d^4 l}{(2 \pi)^4}~
\frac{l^\nu (l - k - p - q)^\mu}{l^2 \left[ (l-k-p)^2 - m_c^2 \right] \left[ (l - k - p - q)^2 - m_s^2  \right]} \nonumber\\ 
&=&
\frac{1}{(4 \pi)^2} \int_0^\infty d \alpha d \beta d \gamma~ e^{A+\beta m_c^2 + \gamma m_s^2}
\int \frac{d^4 k}{(2 \pi)^4}~ \frac{k_\mu}{k^2} ~
\Big[
- \frac{g^{\mu \nu}}{2(\alpha+\beta+\gamma)^3} \nonumber\\ && \qquad
+ \frac{\left[ (\beta+\gamma)k + s \right]^\mu \left[ (\beta+\gamma)k+s \right]^\nu}{(\alpha+\beta+\gamma)^4}
- \frac{(k+p+q)^\mu [(\beta+\gamma)k+s]^\nu}{(\alpha+\beta+\gamma)^3}
\Big] \\
\label{Pinumumu}
\Pi^{\nu \mu \mu}
&=&
\int \frac{d^4 k}{(2 \pi)^4}~ \frac{k_\nu}{k^2}
\int \frac{d^4 l}{(2 \pi)^4}~
\frac{l^\mu (l - k - p - q)^\mu}{l^2 \left[ (l-k-p)^2 - m_c^2 \right] \left[ (l - k - p - q)^2 - m_s^2  \right]} \nonumber\\ 
&=&
\frac{1}{(4 \pi)^2} \int_0^\infty d \alpha d \beta d \gamma~ e^{A+\beta m_c^2 + \gamma m_s^2}
\int \frac{d^4 k}{(2 \pi)^4}~ \frac{k_\nu}{k^2} ~
\Big[
- \frac{g^{\nu \nu}}{2(\alpha+\beta+\gamma)^3} \nonumber\\ && \qquad
+ \frac{\left[ (\beta+\gamma)k + s \right]^\mu \left[ (\beta+\gamma)k+s \right]^\mu}{(\alpha+\beta+\gamma)^4}
- \frac{(k+p+q)^\mu [(\beta+\gamma)k+s]^\mu}{(\alpha+\beta+\gamma)^3}
\Big] 
\end{eqnarray}
Multiplying Eq.~(\ref{Pimumunu}), (\ref{Pimunumu}) and (\ref{Pinumumu}) with the pion four momentum, $q_\mu$, we get
\begin{eqnarray}
\label{qPimumunu}
q_\nu \Pi^{\mu \mu \nu}
&=&
\frac{1}{(4 \pi)^2} \int_0^1 d \gamma d \rho \int_0^\infty d \kappa~ e^{\kappa \rho {\mathcal{Z}}^2}~\Big[
- \frac{\rho}{2 \kappa} \int \frac{d^4 k}{(2 \pi)^4}~e^A~ \frac{q.k}{k^2} \nonumber\\&&
\qquad
+ \rho^3 \int \frac{d^4 k}{(2 \pi)^4}~e^A~ \frac{k.(k+p+\gamma q)~q.(k+p)}{k^2} \nonumber\\&& \qquad \qquad
-\rho^2 \int \frac{d^4 k}{(2 \pi)^4}~e^A~\frac{k.(k+p+ \gamma q)~q.(k+p)}{k^2}
\Big], \\
\label{qPimunumu}
q_\nu \Pi^{\mu \nu \mu}
&=&
\frac{1}{(4 \pi)^2} \int_0^1 d \gamma d \rho \int_0^\infty d \kappa~ e^{\kappa \rho {\mathcal{Z}}^2}~\Big[
- \frac{\rho}{2 \kappa} \int \frac{d^4 k}{(2 \pi)^4}~e^A~ \frac{q.k}{k^2} \nonumber\\&&
\qquad
+ \rho^3 \int \frac{d^4 k}{(2 \pi)^4}~e^A~ \frac{k.(k+p+\gamma q)~q.(k+p)}{k^2} \nonumber\\&& \qquad \qquad
-\rho^2 \int \frac{d^4 k}{(2 \pi)^4}~e^A~\frac{k.(k+ p+ q)~q.(k+p)}{k^2}
\Big],\\
\label{qPinumumu}
q_\nu \Pi^{\nu \mu \mu}
&=&
\frac{1}{(4 \pi)^2} \int_0^1 d \gamma d \rho \int_0^\infty d \kappa~ e^{\kappa \rho {\mathcal{Z}}^2}~\Big[
- \frac{2 \rho}{\kappa} \int \frac{d^4 k}{(2 \pi)^4}~e^A~ \frac{q.k}{k^2} \nonumber\\&&
\qquad
+ \rho^3 \int \frac{d^4 k}{(2 \pi)^4}~e^A~ \frac{k.q~(k+p+\gamma q)^2}{k^2} \nonumber\\&& \qquad \qquad
-\rho^2 \int \frac{d^4 k}{(2 \pi)^4}~e^A~\frac{k.q~(k+p+ \gamma q).(k+p+q)}{k^2}
\Big]
\end{eqnarray}
Eq.~(\ref{qPimumunu}), (\ref{qPimunumu}) and (\ref{qPinumumu}) and (\ref{Pi3c}) allows us to write
\begin{eqnarray}
\Pi_{3Q}
&=&
\frac{1}{(4 \pi)^2} \int d \gamma d \rho \int_0^\infty d \kappa ~ e^{\kappa \rho {\mathcal{Z}}^2} \Big[
\frac{\rho}{\kappa} \int \frac{d^4 k}{(2 \pi)^4} e^A~\frac{k.q}{k^2} \nonumber\\&&
+ 2 \rho^3 \int \frac{d^4 k}{(2 \pi)^4} e^A~ \frac{k.(k+p+q)~q.(k+p)}{k^2} \nonumber\\&&
-\rho^3 \int \frac{d^4 k}{(2 \pi)^4} e^A~ \frac{(q.k)(k+p+\gamma q)^2}{k^2} \nonumber\\&&
-\rho^2 \int \frac{d^4 k}{(2 \pi)^4} e^A~\frac{q.(k+p)}{k^2} \Big\{  
2k^2 +(p+q).k + (p+\gamma q).k \Big\} \nonumber\\&&
+ \rho^2 \int \frac{d^4 k}{(2 \pi)^4} e^A~ \frac{q.k}{k^2} \Big\{
k^2 + k.\big[ (p+q)+(p+\gamma q) \big] + (p+q).(p+\gamma q)
\Big\}
\Big]
\end{eqnarray}
After a little manipulation, we get
\begin{eqnarray}
\Pi_{3Q}
&=&
\frac{1}{(4 \pi)^2} \int_0^1 d \gamma d \rho \int d \kappa~ e^{\kappa \rho {\mathcal{Z}}^2}
\Big[
\frac{\rho}{\kappa} \int \frac{d^4 k}{(2 \pi)^4} e^A~ \frac{k.q}{k^2} \nonumber\\&&
+ 2 \rho^3 \Big\{
\int \frac{d^4 k}{(2 \pi)^4} e^A + (p+ \gamma q)_\mu\int \frac{d^4 k}{(2 \pi)^4} e^A \frac{k^\mu}{k^2}
\Big\}q.(k+p) \nonumber\\&&
- \rho^3 \int \frac{d^4 k}{(2 \pi)^4} e^A~ \frac{(k.q)\left[ k^2 + 2k.(p+ \gamma q) + (p+ \gamma q)^2 \right]}{k^2} \nonumber\\&&
- 2 \rho^2 \int \frac{d^4 k}{(2 \pi)^4} e^A~ q.(k+p) - \rho^2 \int \frac{d^4 k}{(2 \pi)^4} e^A~ \frac{q.(k+p)~k.\Big\{ (p+q)+ (p+\gamma q) \Big\}}{k^2} \nonumber\\&&
+ \rho^2 \int \frac{d^4 k}{(2 \pi)^4} e^A~ q.k  + \rho^2 \int \frac{d^4 k}{(2 \pi)^4} e^A~ \frac{(q.k)~k.\Big\{ (p+q)+(p+\gamma q) \Big\}}{k^2} \nonumber\\&& \qquad
+ \rho^2 \int \frac{d^4 k}{(2 \pi)^4} e^A~\frac{(q.k)(p+q).(p+\gamma q)}{k^2}
\Big]
\end{eqnarray}
we get
\begin{eqnarray}
\Pi_{3Q} &=&
\frac{1}{(4 \pi)^2}
\int_0^1 d \gamma d \rho \int_0^\infty d \kappa~ e^{\kappa \rho {\mathcal{Z}}^2} \Big[
\int \frac{d^4 k}{(2 \pi)^4} e^A~ \frac{k.q}{k^2} \Big\{
\frac{\rho}{\kappa} - (\gamma + 1)\rho^2 q.p \nonumber\\&&
+ 2 \rho^3 \gamma (p.q) - \rho^2 (p+ \gamma q).\left[ \rho(p+ \gamma q) - (p+q)  \right]
\Big\} \nonumber\\&&
+ 2 \rho^2 (\rho -1)(p.q)\int \frac{d^4 k}{(2 \pi)^4} e^A~\frac{k.p}{k^2} + 2 \rho^2(\rho -1)(p.q)\int \frac{d^4 k}{(2 \pi)^4} e^A + \rho^2(\rho -1) \int \frac{d^4 k}{(2 \pi)^4} e^A~ k.q
\Big] \nonumber\\
\end{eqnarray}
But
\begin{eqnarray}
- (\gamma + 1)\rho^2 q.p + 2 \rho^3 \gamma (p.q) - \rho^2 (p+ \gamma q).\left[ \rho(p+ \gamma q) - (p+q)  \right]
&=& - \rho^2 (\rho - 1) p^2
\end{eqnarray}
This eventually gives us
\begin{eqnarray}
\label{Pi3c}
\Pi_{3Q}
&=&
\frac{1}{(4 \pi)^2} \int d \gamma d \rho~ \rho^2 (\rho -1) \int_0^\infty d \kappa~ e^{\kappa \rho {\mathcal{Z}}^2} \Big[
-p^2 \int \frac{d^4 k}{(2 \pi)^4} e^A~\frac{k.q}{k^2} + 2p.q \int \frac{d^4 k}{(2 \pi)^4} e^A~\frac{k.p}{k^2} \nonumber\\&& \qquad
+2p.q \int \frac{d^4 k}{(2 \pi)^4} e^A + \int \frac{d^4 k}{(2 \pi)^4} e^A~(k.q) + \frac{\rho}{\kappa} \int \frac{d^4 k}{(2 \pi)^4} e^A~\frac{k.q}{k^2}
\Big]
\end{eqnarray}
Using 
\begin{eqnarray}
\int \frac{d^4 k}{(2\pi)^4} e^A &=&\frac{1}{2(4 \pi)^2} \frac{e^{f - F^2/4D}}{D^3}~(2D),\\
\int \frac{d^4 k}{(2\pi)^4} ~k^\mu~ e^A &=& 
\frac{1}{2(4 \pi)^2} \frac{e^{f - F^2/4D}}{D^3}~(-F^\mu),\\
\int \frac{d^4 k}{(2\pi)^4} ~\frac{k^\mu}{k^2}~ e^A &=& \int_0^\infty d \lambda \frac{e^{f - F^2/4(D- \lambda)}}{2(4 \pi)^2(D-\lambda)^3} \left( -F^\mu \right), \\
\int \frac{d^4 k}{(2\pi)^4} ~\frac{k^\mu k^\nu}{k^2}~ e^A &=&
\int_0^\infty d \lambda \frac{e^{f - F^2/4(D-\lambda)}}{2(4 \pi)^2(D-\lambda)^3} \left[
g^{\mu \nu} + \frac{F^\mu F^\nu}{2(D- \lambda)}
\right]
\end{eqnarray}
we get
\begin{eqnarray}
\Pi_{3Q}
&=&
\frac{1}{2(4 \pi)^4}
\int_0^1 d \gamma d \rho 2 \rho^3 (1 - \rho)^2~(p.q) \int_0^1 d \kappa \frac{1}{g(\rho, \kappa, \gamma)^3}\Big[
\kappa (p+ \gamma q)^2 \Gamma(\epsilon) a^{-\epsilon} \nonumber\\&&
\qquad \qquad + \rho \Gamma(-1 + \epsilon)a^{1 - \epsilon}
\Big]
+ \frac{1}{2(4 \pi)^4} \int_0^1 d \gamma d \rho~ \frac{2p.q}{(1- \rho)}~ \left[ \Gamma(-1 + \epsilon)(a')^{1 - \epsilon} \right]
\end{eqnarray}
where
\begin{eqnarray}
{\mathcal{Z}}^2(\gamma) &=& (1 - \gamma)m_c^2 + \gamma m_s^2, \\
g(\rho, \kappa) &=&\kappa \rho(1 - \rho)+ (1 - \kappa), \\
a(\rho, \kappa, \gamma) &=& - \kappa \rho {\mathcal{Z}}^2 + \kappa \rho (1 - \rho) p.(p+ \gamma q) - \Big\{ \frac{\kappa \rho (1 - \rho)}{g} \Big\} \kappa \rho (1 - \rho) (p+ \gamma q)^2, \\
a'(\rho, \gamma) &=& - \rho {\mathcal{Z}}^2 + \rho (1 - \rho) p.(p+ \gamma q) -
 \rho (1 - \rho) (p + \gamma q)^2 ~.
\end{eqnarray}
\end{appendix}

\end{document}